\documentclass{aa}  
\usepackage{graphicx}
\usepackage{chemformula} 
\usepackage[T1]{fontenc} 
\usepackage{longtable}
\usepackage{adjustbox}
\usepackage{lscape}
\usepackage{hyperref}
\usepackage{nameref}
\usepackage{txfonts}
\usepackage{xcolor}
\usepackage{subfig}
\usepackage{siunitx}
\usepackage{multirow}
\usepackage{tabularx,booktabs,ragged2e}

\begin{document}

   \title{Predicting binding energies of astrochemically relevant molecules via machine learning}
   \titlerunning{Predicting binding energies of astrochemical molecules via machine learning}
   
   \author{T. Villadsen
          \inst{1},
          N.F.W. Ligterink
          \inst{2}
          \and
          M. Andersen
          \inst{1,}
          \inst{3}
          }

   \institute{Department of Physics and Astronomy - Center for Interstellar Catalysis, Aarhus University, Aarhus C, DK‐8000 Denmark\\
              \email{mie@phys.au.dk}
         \and
             Physics Institute, University of Bern, Sidlerstrasse 5, 3012 Bern, Switzerland\\
             \email{niels.ligterink@unibe.ch}
        \and
         Aarhus Institute of Advanced Studies, Aarhus University, Aarhus C, DK‐8000 Denmark
        }

\date{Received May 23, 2022; accepted }

 
  \abstract
   {The behaviour of molecules in space is to a large extent governed by where they freeze out or sublimate. The molecular binding energy is thus an important parameter for many astrochemical studies. This parameter is usually determined with time-consuming experiments, computationally expensive quantum chemical calculations, or the inexpensive, but inaccurate, linear addition method. 
   } 
   {In this work we propose a new method based on machine learning for predicting binding energies that is accurate, yet computationally inexpensive.
   }
   {A machine learning model based on Gaussian Process Regression is created and trained on a database of binding energies of molecules collected from laboratory experiments presented in the literature. The molecules in the database are categorized by their features, such as mono- or multilayer coverage, binding surface, functional groups, valence electrons, and H-bond acceptors and donors.}
   {The performance of the model is assessed with five-fold and leave-one-molecule-out cross validation. Predictions are generally accurate, with differences between predicted and literature binding energies values of less than $\pm$20\%. The validated model is used to predict the binding energies of twenty one molecules that have recently been detected in the interstellar medium, but for which binding energy values are not known. A simplified model is used to visualize where the snowlines of these molecules would be located in a protoplanetary disk.}
   {This work demonstrates that machine learning can be employed to accurately and rapidly predict binding energies of molecules. Machine learning complements current laboratory experiments and quantum chemical computational studies. The predicted binding energies will find use in the modelling of astrochemical and planet-forming environments.}


   \maketitle
%

\section{Introduction}

    One of the objectives of the field of astrochemistry is to understand the formation, destruction, and survival of molecules in astrophysical environments, such as star-forming regions or planet-forming disks \citep{jorgensen2020,oberg2021}. This concerns, for instance, the molecular reactions on dust-grain surfaces, where the surface acts as a catalyst \citep{cuppen2017} and evaporation of ices in hot cores near young massive stars \citep{viti2004}. To model such phenomena certain molecule-specific parameters are needed. Both examples require a measure of how strongly the molecule binds to the surface, that is the binding energy (BE). In astrophysical environments physisorption is the primary source of the BE. The main contributors of physisorption are van der Waals forces, which arises from dipole-dipole interactions, and hydrogen bonding between the adsorbed molecule and surface. These forces act at a distance between 2-4 \si{\angstrom}. If no dynamical barrier is present, the BE is equal to the activation energy of desorption ($E_{\mathrm{des}}$) as the system reference energy (i.e., $E$ = 0) corresponds to the adsorbed molecule and surface at infinite separation \cite{minissale2022}. 
    
    Experimentally, a prominent technique to determine the BE is by temperature programmed desorption (TPD). This applies to studies of catalysis \citep{Luo1997}, surface science \citep{Zhou2007}, as well as astrochemistry \citep{Caro2010}. The TPD process consists of three steps; first, the molecule is adsorbed to the surface at cold temperatures. The coverage may be below or above a single monolayer depending on various factors such as the molecular flux and deposition time. If more molecules than available surface adsorption sites are deposited, the coverage is also termed multilayer. Second, the temperature is linearly increased, resulting in desorption of molecules at specific temperatures. Third, the desorbed molecules are detected, often with a mass spectrometer. This produces spectra of desorption rate as a function of temperature. The process of thermal desorption is generally described by the Polanyi-Wigner equation, a modified Arrhenius law:
    
    \begin{equation}
        -\frac{d N}{d t}  = k_{\rm des}(T) \cdot N^{\rm n},
    \end{equation}
    
    \begin{equation}
        -\frac{d N}{d t}  = \nu_{\rm n} \cdot N ^{\rm n} \cdot {\rm exp} \left( -\frac{E_{\rm des}}{k_{\rm B} \cdot T} \right),
        \label{eq:polanyi}
    \end{equation}
    where ~$k_{\rm des}(T)$ is the desorption rate constant in s$^{-1}$ at temperature $T$, $N$ the number of adsorbed molecules on a surface, $n$ the order of desorption (usually 1 for monolayer desorption and 0 for multilayer desorption), $\nu_{\rm n}$ the pre-exponential frequency factor with value molecules$^{\rm 1-n}$ s$^{-1}$ (also often denoted as $A$), $E_{\rm des}$ the desorption energy and $k_{\rm B}$ the Boltzmann constant. For a more thorough discussion of the technique and analysis, we refer to \citet{dejong1990}, \citet{burke2010}, and \citet{minissale2022} for contextual reviews.
   
    While TPD has been successful in determining the BEs for many molecules that are of astrochemical relevance \citep[e.g.,][]{brown2007,burke2015a,smith2018,behmard2019,salter2019}, there are limitations to experimental BE investigations with this and other techniques. Experiments are time-consuming and the focus usually is on the scientifically most impactful systems. With the vast number of known interstellar molecules, this inevitably means that some of them are not yet studied. Furthermore, certain molecular species are difficult to work with, either because they are unstable or highly reactive (e.g., vinyl alcohol, cyanopolyynes), highly toxic (e.g., methyl isocyanate, propyl cyanide), or simply challenging to produce an ice film with (e.g, carbamide). 
    
    Alternatively, Bayesian inference \citep{heyl2022} or quantum chemical computational methods can be used to determine BEs \citep[e.g.,][]{das2018,rimola2018,balbisi2022}.
    Quantum chemical calculation can also take into account BE distributions on amorphous and highly anisotropic surfaces \citep[e.g.,][]{tinacci2022, ferrero2020, duflot2021}.
    However, since these methods are computationally expensive, many astrochemical studies rely on the so-called linear addition method. With this method, the BE of a molecule is determined by splitting its components in atoms and molecular fragments for which the BEs are known and subsequently adding them together \citep[e.g.,][]{garrod2006,shingledecker2020}. This method is computationally inexpensive, but it is also inaccurate. Novel methods that are computationally inexpensive, but more accurate are required. 
    
    Machine learning (ML) has become one of the most prominent scientific tools of the $21^{\mathrm{st}}$ century as it provides high accuracy at low computational cost. It has the ability to handle and interpret data in ways impossible to humans, which allows for the discovery of unprecedented patterns \citep{Jordan2015}. These properties make ML an interesting alternative to the above-mentioned theory-based approaches. With immense versatility ML has applications ranging from self-driving cars and social media to banking and image recognition. In recent years, ML has also made its entrance as a powerful tool in astrochemistry and astrophysics. Notable deployments include \cite{lee2021} for reproducing and predicting chemical abundances in interstellar inventories and \cite{Shallue2018} for exoplanet identification.
    Another type of ML models are ML interatomic potentials, which can be directly employed as low-cost alternatives to quantum chemical calculations for investigating for example molecular reactivity, adsorption and diffusion on dust grains \citep[e.g.,][]{sevillano2021, molpeceres2021, zaverkin2021}. In surface science and catalysis, ML has also been used extensively to identify predictive models for BEs \citep[e.g.,][]{Gu2020,Fung2021,andersen2021adsorption}.
    
    In this work, we apply supervised ML to a data set of BEs obtained from literature TPD data and thereby develop a model to predict BEs between new molecules and surfaces relevant to astrochemical environments. The methods used are discussed in Sect.\ 2. The results and discussion of the analysis are presented in Sect.\ 3. Astrophysical implications are covered in Sect.\ 4 and the conclusions of this work are given in Sect. 5.

\section{Methods and data}

\begin{figure*}
    \centering
    \includegraphics[width=\hsize]{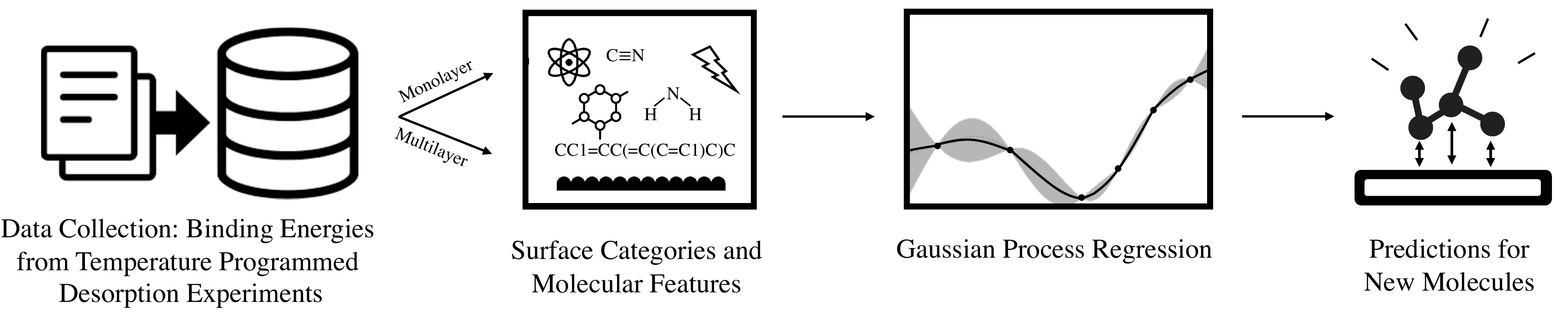}
    \caption{Schematic overview of the workflow. First the data is collected from the literature and divided into monolayer and multilayer coverage. Secondly, specific features are designated to the data, including atomic composition, functional groups and valence electrons. Thirdly, the Gaussian Process Regression model is constructed and trained on the data to be able to predict BEs for new molecules.}
    \label{workflow}
\end{figure*}

Supervised learning algorithms are constructed to make a model that can recognise particular patterns within the data when given training examples by the user (supervisor). A strong limitation of such algorithms is that they can only recognise data that are related to the training data, therefore, any anomaly or unseen data structures would be difficult for the model to grasp. The training data given to the model in our work is a data set of BEs obtained from TPD experiments collected from the literature as well as relevant features of each system such as the surface category and atoms and functional groups present in the adsorbed molecule. Hence, the trained model can be expected to predict BEs for new examples of molecules and surfaces that are not too different from those seen in the training data. We quantify the predictive accuracy of our model by carrying out two types of cross validation analysis. The workflow of the process is shown in Fig. \ref{workflow}. In the following sections, the essential components of this workflow are described.
    
\subsection{Gaussian process regression}
\label{sec:GPR}
BEs are here predicted using the supervised ML technique Gaussian Process Regression (GPR). It is a probabilistic, non-parametric supervised learning method frequently used for regression and classification problems in the ML community. Being based on Bayesian probability theory, it learns a posterior probability distribution over all admissible target functions. Here, these are functions describing the relationship between surface/molecular features (the input $x$) and the TPD BE (the output $y$).
A Gaussian Process prior is assumed, which means that both the prior and posterior probability distributions are Gaussian distributed (normal) and can be specified using a mean function, $\mu(x)$, and a covariance function, $k(x,x')$, also called the kernel function. The posterior distribution is calculated by conditioning the prior distribution on the training data set. Model predictions on new test data points with input $x_*$ are obtained from the mean of the posterior distribution, $\mathbf{\bar{f}}_*$, given by
\begin{equation}\label{eq:gpr_mean}
    \mathbf{\bar{f}}_* = \boldsymbol{\mu_*} + k(x_*,x)[k(x,x) + {\sigma_{n}^2} I]^{-1}(\mathbf{y}-\boldsymbol{\mu}) \quad , 
\end{equation} 
    and variances are obtained from the diagonal of the covariance matrix, ${\rm cov}(\mathbf{f}_*)$, given by
\begin{equation}\label{eq:gpr_cov}
    {\rm cov}(\mathbf{f}_*) = k(x_*,x_*)-k(x_*,x)[k(x,x) + {\sigma_{n}^2} I]^{-1}k(x,x_*) \quad .
\end{equation}
Here $\boldsymbol{\mu}$ and $\boldsymbol{\mu_*}$ are the mean vectors, $k(x,x_*)$ denotes the covariance matrix evaluated at all pairs of training and test points, and similarly for the other entries $k(x,x)$, $k(x_*,x_*)$ and $k(x_*,x)$. The target function is assumed to be noisy, which is accounted for by the incorporation of independently, identically distributed Gaussian noise, $\sigma_{n}^2I$. In practise, a small or vanishing noise level will cause the fitted model to follow the training data points closely, whereas a higher noise level will result in a smoother model. The latter can be useful for extrapolating to unseen data since more emphasis is put on trends rather than the individual training examples seen. The variance provides an uncertainty estimate, that is how confident we can be about the model predictions, and it is also affected by the noise level. The direct access to an uncertainty estimate is a great advantage of GPR compared to other types of ML methods such as neural networks \citep{scalia2020}.
        
GPR belongs to the class of kernelized ML methods that employ internally the “kernel trick”. A kernel is a function that corresponds to an inner product in some high-dimensional feature space, whose values can be interpreted as a similarity measure of input data points. In this way, the kernel function implicitly maps the inputs into a higher dimensional space and applies the linear algorithm there, which is the basic idea how kernelized ML methods tackle the computational complexity of dealing with high-dimensional feature spaces. We refer to \citet{rasmussen2006} for an extensive textbook coverage of GPR and to \citet{gibson2012}) and \citet{aigrain2012} for examples of GPR applied to exoplanet data sets.
    
    For the actual GPR implementation, we rely in this work on the ML library Scikit-learn \citep{pedregosa2011}, which provides a number of built-in kernels. By testing several different combinations of kernels using five-fold cross validation (cf.\ Sect.\ \ref{sec:5fCV}), we found that the best performance is achieved by using the sum of the radial basis function (RBF) kernel
    
    \begin{equation}
        k_{\mathrm{RBF}}(\textbf{x}_i,\textbf{x}_j) = \exp{\left[ -\frac{1}{2\ell_1^2} (\textbf{x}_i-\textbf{x}_j)^2 \right]}
    \end{equation}
    and the rational quadratic (RQ) kernel
    \begin{equation}
        k_{\mathrm{RQ}}(\textbf{x}_i,\textbf{x}_j) = \exp{\left[ 1 + \frac{(\textbf{x}_i-\textbf{x}_j)^2}{2 \alpha \ell_2^2}  \right]}^{-\alpha} \quad .
    \end{equation}
    The length-scale parameters, $\ell_1$ and $\ell_2$, indicate how quickly the correlation between two points drops as their distance increases. A higher $\ell$ gives a smoother function and a smaller $\ell$ gives a wigglier function \citep{rasmussen2006}. $\alpha$ determines the relative weighting of large-scale and small-scale variations in the RQ kernel \citep{duvenaud2014}. $\ell_1$, $\ell_2$ and $\alpha$, along with the noise level from Eq.\ \ref{eq:gpr_mean} and \ref{eq:gpr_cov}, are hyperparameters. Here we determine these parameters during the model training by maximising the marginal likelihood function using the standard Limited-memory Broyden-Fletcher-Goldfarb-Shanno algorithm for bound-constrained optimization (L-BFGS-B) \cite{byrd1995limited,zhu1997algorithm}. 
    
    
    \subsection{Data set and features}
    \label{sec: data}
    We have compiled the training data by analysing laboratory studies presented in the literature and extracting the relevant information. From these publications, BEs determined with TPD experiments and information about the binding surface are retrieved, which resulted in a data set that initially contained 354 (167) entries for the monolayer (multilayer) case and 117 different molecules, many of which have been detected in the interstellar medium. They range from simple diatomics like N$_{2}$ and CO, organic molecules like ethanol (CH$_{3}$CH$_{2}$OH) and glycolaldehyde (HOCH$_{2}$CHO), long carbon chains such as octane (C$_{8}$H$_{18}$), and biomolecules like the nucleobase adenine. 

    Besides the BEs, the training data set also contains input features, which are descriptive attributes of the surfaces and molecules. It is essential to choose the best possible features as they govern how well the model predicts; too few features and the model will not be able to differentiate between different training data points (e.g.\ molecules), and too many features could invoke the curse of dimensionality, which is a term expressing how increasing the volume of feature space dilutes the data \citep{bellman1966}. Therefore, the best approach is to minimise the number of features while maximising the amount of information they contain. 
    
    In this work, molecular features, such as atomic compositions (C, H, O, N, Cl) and functional groups (alcohol, --OH; carbonyl, --C(O)--; carboxyl, --COOH; ester, --C(O)O--;  ether, --O--; amine, --NH$_{2}$; cyanide, --CN; amide, --NC(O)--), are used. Several features are obtained from the python module RDkit \citep{landrum2020}. These are calculated by converting the molecules to SMILES\footnote{Simplified Molecular-Input Line-Entry System (SMILES) is a formalism widely used in the chemistry community to describe molecular structures as a string of ASCII characters. The SMILES string also encodes which chemical functional groups are present in the molecule.} 
    strings, which are then fed into RDkit. The considered features are the number of valence electrons, hydrogen bond donors, hydrogen bond acceptors, and topological polar surface area (TPSA). The latter is a property defined as the molecule's sum of surface area of polar atoms and is measured in Å$^2$. The motivation for applying the features from RDkit is to try to encapsulate the origin of the dominating forces that govern the binding of molecules. This includes in particular hydrogen bonding and van der Waals interactions. Finally, we also included the molecular mass and the number of atoms in the molecule as features. An overview of all molecular features used is given in Table \ref{tab:features} and a selection of the feature values of each molecule are presented in Table \ref{tab:molecule_features}. 
    
    \begin{table*}
    \caption{Overview of the features used to describe the molecules and surfaces.}
    \centering
    \begin{tabular}{l l l l l}
    \hline \hline
    Atoms & Functional Groups & RDKit \& Misc. & Surface & Examples of surface\\
    \hline
    Carbon   & Alcohol \phantom{i} (--OH) & Number of H-bond acceptors & Carbon & Graphene, graphite
    and \\
    Chlorine & Amide \phantom{iii} (--NC(O)--)   & Number of H-bond donors &   & highly orientated pyrolytic graphite \\
    Hydrogen & Amine \phantom{iii} (--NH$_2$)   & Number of valence electrons &  Metal & Gold and nickel \\
    Nitrogen & Carbonyl (--C(O)--) & Topological polar surface area  &   \\
    Oxygen   & Carboxyl (--COOH)  &            & Silicate & Amorphous silicate and forsterite\\
             & Cyanide \phantom{|} (--CN)      & Mass      &  \\
             & Ester \phantom{iiiii} (--C(O)O--)   & Number of atoms    & Water & Amorphous solid water and \\
             & Ether \phantom{i||ii} (--O--)       &     & & crystalline water\\
    \hline
    \hline
    \label{tab:features}
    \end{tabular}\
    \end{table*}

For the consideration of features related to the surface, we first divide our data set into two categories; monolayer and multilayer coverage. Entries recorded at monolayer coverage (that is, an adsorbate layer of one molecule thickness) or less are assumed to have their BE dominated by the surface-molecule interaction. A wide variety of different surfaces are present in the analysed literature, the number of which greatly exceeds what is reasonable for the model to handle. To reduce this, the surfaces are placed in four different sub-categories based on their common traits. The categories and their main contributors annotated with the percentage they comprise of the total entries in the category are the following: Carbon (graphene, graphite and highly orientated pyrolytic graphite, 94 \% of data entries), metal (gold, 56 \% of data entries), silicate (amorphous silicate and forsterite, 93 \% of data entries) and water (amorphous solid water and crystalline water, 98 \% of data entries). We note that this approximation could be a source of significant noise in the training data, as different types of surfaces here placed in the same category (e.g.\ nickel and gold for metals) may in reality bind the studied molecules with different strength, however, they cannot be distinguished from the used input features. Finally, in order to input the surface category feature to the ML algorithm it is converted to numerical values using one hot encoding. 
    
For multilayer entries, the coverage is greater than one molecule in thickness and the BE is assumed to be dominated by intermolecular interactions of the adsorbed species with itself. For this reason, we have chosen to neglect surface features for the multilayer data set. 
    
Finally, we note that, while for further analysis only the BE is used, it is important to be aware of the influence of the pre-exponential factor. This value can be experimentally determined, but often (including in many of the studies that contribute to the data set in this work) an assumed value is used in the analysis to retrieve the BE. Compared to an experimentally determined pre-exponential factor, the assumed value may result in significantly deviating BEs and their inclusion in the training data will affect the results of the ML model. It is also worth noting that adsorbates on amorphous and highly anisotropic surfaces usually have a distribution of BEs rather than a single value \citep{shimonishi2018}. However, since the available TPD data mainly consists of a single binding energy per surface/molecule combination, this distribution is presently not possible to include in our model.

\subsection{Data preparation}

Since ML algorithms generally do not perform well on data points that are rare or very different from the remaining of the data set, we combed our data set for outliers. This was done by applying the `isolation forest algorithm' from Scikit-learn \citep{pedregosa2011}. It identifies anomalies that are both few in numbers and different in feature space. The isolation forest has been applied in both the mono- and multilayer case. It resulted in removal of three outliers for both cases: the C$_{60}$ fullerene and the two polycyclic aromatic hydrocarbons (PAHs) coronene and ovalene. For the multilayer case we also removed two other molecules (dotriacontane and guanine) since these molecules, together with the fullerene and the PAHs, are outliers in the sense that they have BEs that are 3-4 times as large as the other molecules in the data set.

Lastly, we addressed the issue that data points with the same feature representation can have different labels (BEs). This can arise when several different experimental measurements of the same surface/molecule combination are present in the data set, which differ only in experimental parameters that in our model have been assumed not to influence the BE, such as temperature ramp, starting temperature and pre-exponential factor. In this case an average over the measured BEs are used. If, on the other hand, this issue arises due to differences in parameters such as the initial coverage and the specific surface (for the monolayer data set), a more thoughtful approach is fruitful. If data points at both sub-monolayer and monolayer coverage are present, we use the monolayer data point since the sub-monolayer case may be dominated by specific surface sites exhibiting a more favourable (stronger) binding than the others. If data points for several specific surfaces or facets are present within the same surface category, we use the surface that occurs most frequently within the category in order to get a more coherent data set. After this data preparation step, the final data set contains 143 (46), entries for the monolayer (multilayer) case and 114 individual molecules. After the data preparation step the features were normalised to have zero mean and unit variance, with the exception of the one-hot-encoded surface features. This is a standard procedure in the ML community to make the learning task easier.
All data points used in this work, their actual surface and assigned surface category are given in Table \ref{tab:monolayer_data} and \ref{tab:multilayer_data} for monolayer and multilayer coverage species, respectively\footnote{Electronic versions of the data files, along with Python scripts for producing the results presented below, can be found in the Github repository \href{https://github.com/TorbenVilladsen/Predicting-Binding-Energies-of-Astrochemically-Relevant-Molecules-via-Machine-Learning.git}{here.}}.
    

\section{Results and discussion}
In the following sections we validate the performance of the model using two different types of assessment; five-fold cross validation and leave-one-molecule-out cross validation.

\subsection{Five-fold cross validation}
\label{sec:5fCV}
\begin{figure*}[hbt]
        \centering
        \subfloat[Monolayer \label{PPmono}]{\includegraphics[scale=0.42]{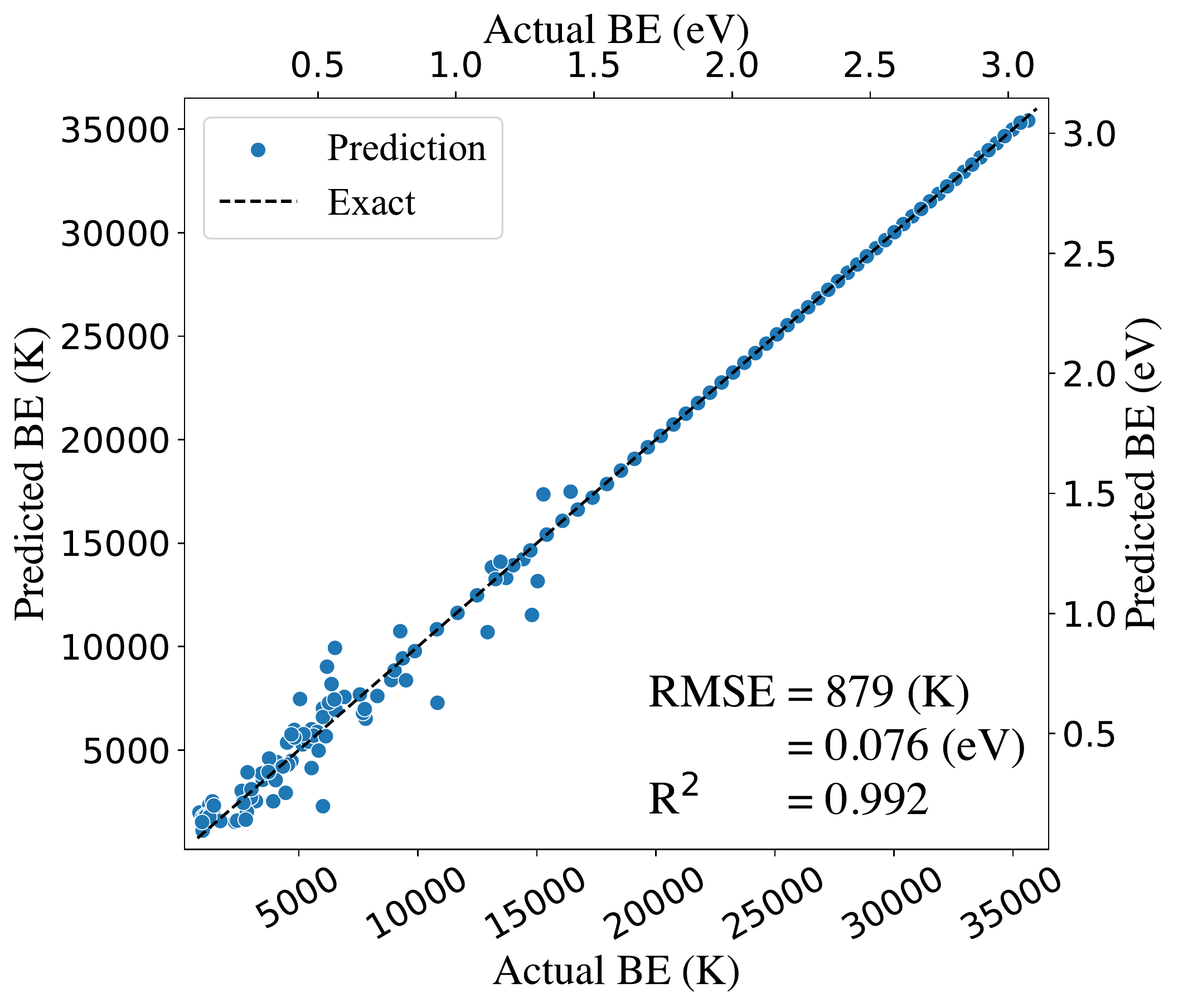}}
        \quad
        \subfloat[Multilayer \label{PPmulti}] {\includegraphics[scale=0.42]{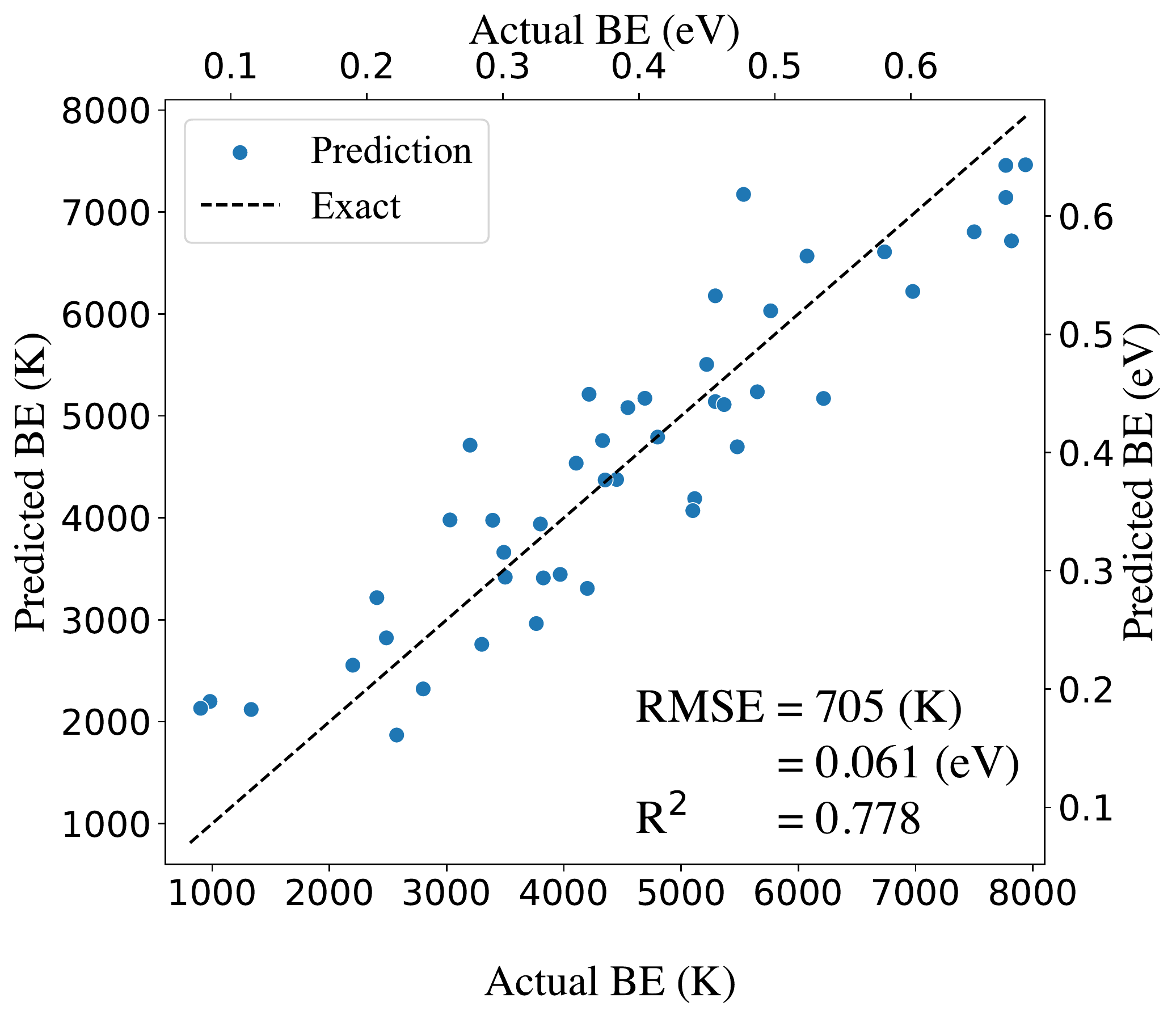}}
        \caption{Parity plots for a) monolayer and b) multilayer coverage comparing ML-predicted BEs against actual BEs for the combined validation set from five-fold cross validation.}
        \label{PP}
    \end{figure*}

Five-fold cross validation is a standard approach in the ML community for comparing models and for assessing how well they can predict new data points. As illustrated in Fig. \ref{crossval} in Appendix \ref{ap:cross}, it is carried out by splitting the data set into five disjoint equally sized sets. The model is then trained on four of the sets while the fifth set is used to validate the model (i.e.\ by comparing model predicted BEs to actual BEs). This is repeated five times until every data point has been used as validation data exactly once. In Fig.\ \ref{PP} we show parity plots of ML predictions on the combined validation data set from the five folds versus actual literature BEs for the monolayer and multilayer data sets. The closer the points are to the diagonal dotted line, the better the model performs. The performance is quantified using the root-mean-square error (RMSE) and more absolutely by the coefficient of determination, $R^2$.  

It is found that the model has the highest $R^2$ value for the monolayer case shown in Fig. \ref{PPmono}. One contributor to this could be the fact that there are more than three times as many entries for the monolayer set than for the multilayer set. However, a much more important difference between the two data sets is that for the monolayer case the model is exceptionally accurate for data points with a BE above 17000~K. The reason for this behaviour is that this part of the data set is immensely uniform as it consists only of carbohydrate chains of varying lengths adsorbed on graphene surfaces. Since the BE increases proportionally with the length of the carbon chain, the model is able to learn these BEs with a very high accuracy. In fact, if only the entries with BE above 17000~K are included in the analysis, the model achieves a RMSE of 25.7 K and an $R^2$ value of approximately one. This RMSE is much lower than the overall RMSE of 879 K, especially when taken into consideration that the scale of the data set is higher. However, it should be kept in mind that the model is only as nuanced as the data it has been provided. This implies that the model has a very narrow knowledge of molecules in the high BE regime, and thus the predictive capability would most likely be limited for molecules different from the simple carbohydrates. If only the low energy regime is considered, the $R^2$ value decreases to 0.946. The corresponding parity plot, Fig. \ref{PPlow}, can be found in the appendix.
    
\subsection{Leave-one-molecule-out cross validation}
As a second type of assessment, we next evaluate how well the model can predict individual new molecules. This is done by first removing a chosen molecule from the data set (including all surface categories for the monolayer case), then training the model on the reduced data set, and finally comparing the ML-predicted BE of the chosen molecule with the literature value. For this assessment eight different molecules have been chosen, namely acetone ($\mathrm{CH}_3 \mathrm{C(O)CH}_3$), acetonitrile ($\mathrm{CH}_3 \mathrm{CN}$), allyl alcohol ($\mathrm{C}_3 \mathrm{H}_5 \mathrm{OH}$), ammonia ($\mathrm{NH}_3$), methane ($\mathrm{CH}_4$), methylformate ($\mathrm{CH}_3 \mathrm{OCHO}$), the alkane nonacosane ($\mathrm{C}_{29} \mathrm{H}_{60}$) and the nucleobase thymine ($\mathrm{C}_5 \mathrm{H}_6 \mathrm{N}_2 \mathrm{O}_2$), which represent diverse chemical compositions and molecular sizes. 

The comparison between ML-predicted and actual BEs are shown in table \ref{tab:predictnew}. The model uncertainty estimates are obtained from the standard deviation of the posterior distribution (i.e.\ the square root of its variance), as described in Sect.\ \ref{sec:GPR}. As seen, the overall predictive capability of the model is reasonably good. Furthermore, there is a direct correlation between how well different types of molecules are represented in the training data and how well the model predicts them. For example, it is found that nonacosane is exactly predicted, which is presumably a consequence of the many similar molecules in the data set. It is further noticeable that ammonia is predicted quite reasonably, even though the model is mostly trained on organic molecules. We can identify that the good prediction accuracy mainly comes from the inclusion of the following molecular features; number of valence electrons, H-bond acceptors, H-bond donors and TPSA, since the predicted BE for a monolayer of ammonia on a carbon surface is only $2070 \pm 1740$ K (a deviation of $-31$ \% compared to the literature value) if these features are excluded. For the multilayer case the deviation would be even more pronounced at $-36$ \%. The model struggles the most with allyl alcohol, acetontrile and methyl formate, although the deviations are still relatively small at $\pm$ $\sim$20 \%. This might be because the training data set contains few molecules like these three or because another factor such as the specific surface has an influence on the BE, which the model cannot account for.

    \begin{table*}
    \caption{Comparison between ML-predicted and literature BE values (rounded to the nearest ten) 
    for leave-one-molecule-out cross validation.}
    \centering
    \begin{tabular}{l l l c c r}
    \hline \hline
    Name & Molecule & Surface or Coverage & Prediction & Literature & Deviation \\
    & & & (K) & (K) & \\
    \hline
    Acetone & $\mathrm{CH}_3 \mathrm{C(O)CH}_3$ & Water & \phantom{ii} 4520 $\pm$ \phantom{ii}450 \phantom{ii} & \phantom{i} 4690 $\pm$ 240 & $-3.6$ \%  \\
    Acetonitrile & $\mathrm{CH}_3 \mathrm{CN}$ & Metal & \phantom{ii} 6730 $\pm$ \phantom{ii}880 \phantom{ii} & \phantom{i} 5530 $\pm$ 360 & $21$ \%  \\
    Allyl alcohol & $\mathrm{C}_3 \mathrm{H}_5 \mathrm{OH}$ & Metal & \phantom{ii} 7110 $\pm$ 1740 \phantom{ii} & 6010$^{\dagger}$ \phantom{iiiii} & $18$ \%  \\
    Ammonia & $\mathrm{NH}_3$ & Carbon & $2870 \pm 1700$  & \phantom{i} 2990 $\pm$ 240 & $-4.0$ \%  \\
    Methane & $\mathrm{CH}_4$ & Carbon & \phantom{ii} 1870 $\pm$ \phantom{ii}470 \phantom{ii} & \phantom{i} 1700 $\pm$ 120& $10$ \%  \\
    Methyl formate & $\mathrm{CH}_3 \mathrm{OCHO}$ & Water & \phantom{ii} 5390 $\pm$ \phantom{ii}540  \phantom{ii} & \phantom{i} 4510 $\pm$ 530& $19$ \%  \\
    Nonacosane & $\mathrm{C}_{29} \mathrm{H}_{60}$ & Carbon & \phantom{i} 23720 $\pm$ \phantom{i} 220 \phantom{iii} & 23720 $\pm$ 930 & $0.0$ \%   \\
    Thymine & $\mathrm{C}_5 \mathrm{H}_6 \mathrm{N}_2 \mathrm{O}_2$ & Metal & $11680 \pm 2600$ \phantom{i} & 12930 $\pm$ 240 & $-9.6$ \%  \\
     
    \hline
    Acetonitrile & $\mathrm{CH}_3 \mathrm{CN}$ & Multilayer & \phantom{ii} 4030 $\pm$ \phantom{ii}470 \phantom{ii} & \phantom{i} 4800 $\pm$ 190 & $-16$ \%  \\
    Ammonia & $\mathrm{NH}_3$ & Multilayer & \phantom{ii} 3330 $\pm$ \phantom{ii}960 \phantom{ii} & \phantom{i} 3030 $\pm$ 240 & $9.9$ \%  \\
    Methyl formate & $\mathrm{CH}_3 \mathrm{OCHO}$ & Multilayer & \phantom{ii} 4520 $\pm$ \phantom{ii}470 \phantom{ii} & \phantom{i} 4110 $\pm$ 200& $10$ \%  \\
    \hline
    \label{tab:predictnew}
    \end{tabular}\
    \tablefoot{$^{\dagger}$Literature study does not present uncertainty of allyl alcohol BE measurement. A version of this table with the BEs in eV is shown in Table \ref{tab:predictnewEV}}
    \end{table*}

We also recall here that some types of molecules are not well represented in the training data, meaning that the developed model is not suitable for predicting these. This concerns foremost the types of molecules that were removed as outliers in the data preparation step (fullerenes, PAHs, and large aromatic molecules in general). Also phenols are notoriously difficult to perform TPD experiments with, and therefore not much data is available for the model to train on.
    
In general, a further limitation of the developed model is that it cannot distinguish between isomers of the same molecule, in the case these have the same feature representation, which will always be the case for structural isomers. However, given the other uncertainties - both in the experimental TPD procedure and in the developed model - it is very likely that the difference in BE between two isomers would anyway be much smaller than what can be resolved with the current approach. 

Another limitation of the model is that it struggles with molecules consisting of atoms other than C, H, O, N and Cl. This is because the data set includes very few molecules that contain other than said atoms and because we consequently do not include features representing any other atoms. This means that predicting the BEs of for example sulphurous- and phosphorous-containing molecules will be less accurate. To increase the performance and predictive capabilities of the model, the most essential future step would be to increase and diversify the number of entries in the training data set.


\section{Astrophysical implications}
\label{sec:astro}

    \begin{table*}[ht]
    \caption{Predictions of BEs for molecules with astrophysical relevance measured in K and rounded to nearest 10.}
    \centering
    \begin{tabular}{l l l c c c}
    \hline \hline
    Name & Molecule & Surface & Prediction & Prediction & Estimate  \\
    & & & (K) & (eV) & (K) \\
    \hline
    Cyanamide & NH$_{2}$CN & Carbon & $8780 \pm 3730$ & $0.76 \pm 0.32$ & \multirow{2}*{5556}$^{a}$ \\
    Cyanamide & NH$_{2}$CN & Water & $8730 \pm 3710$ & $0.75 \pm 0.32$ &  \\
    Ethanimine & CH$_{3}$CHNH & Carbon & $4350 \pm 1100$ & $0.37 \pm 0.09$ & \multirow{2}*{5580}$^{b}$ \\ 
    Ethanimine & CH$_{3}$CHNH & Water & $4010 \pm 1620$ & $0.35 \pm 0.14$ &  \\
    Vinylalcohol & CH$_2$CHOH & Carbon & 6000 $\pm$ \phantom{i} 260  & $0.52 \pm 0.02$ & -- \\
    Vinylalcohol & CH$_2$CHOH & Water & 5910 $\pm$ \phantom{i} 400 & $0.51 \pm 0.03$ & -- \\
    Propargylimine & HC$_{3}$HNH & Carbon & $4580 \pm 1210$ & $0.40 \pm 0.10$ & \multirow{2}*{14750}$^{c}$ \\ 
    Propargylimine & HC$_{3}$HNH & Water & $4260 \pm 1630$ & $0.37 \pm 0.14$ & \\
    Cyanomethanimine & HNCHCN & Carbon & $6670 \pm 2110$ & $0.57 \pm 0.18$ & \multirow{2}*{10900}$^{c}$ \\ 
    Cyanomethanimine & HNCHCN & Water & $6750 \pm 2100$ & $0.58 \pm 0.18$ &   \\ 
    Methyl isocyanate & CH$_{3}$NCO & Carbon & $3360 \pm 1440$ & $0.29 \pm 0.12$ & \multirow{2}*{6486}$^{d}$ \\
    Methyl isocyanate & CH$_{3}$NCO & Water & $2990 \pm 1490$ & $0.26 \pm 0.13$ &  \\
    Acetamide & CH$_{3}$C(O)NH$_{2}$ & Carbon & 8420 $\pm$ \phantom{i} 520 & $0.73 \pm 0.04$ & -- \\
    Acetamide & CH$_{3}$C(O)NH$_{2}$ & Water & 8350 $\pm$ \phantom{i} 530 & $0.72 \pm 0.04$ &  -- \\
    N-Methylformamide & CH$_{3}$NHCHO & Carbon & $7920 \pm 1290$ & $0.68 \pm 0.11$ & \multirow{2}*{7386}$^{d}$ \\
    N-Methylformamide & CH$_{3}$NHCHO & Water & $7880 \pm 1340$ & $0.68 \pm 0.11$ &  \\
    Carbamide / Urea & NH$_{2}$C(O)NH$_{2}$ & Carbon & $11930 \pm 4350$ \phantom{i} & $1.02 \pm 0.37$ & -- \\
    Carbamide / Urea & NH$_{2}$C(O)NH$_{2}$ & Water & $11960 \pm 4350$ \phantom{i} & $1.03 \pm 0.37$ & -- \\
    Ethenediol & HOCHCHOH & Carbon & $9130 \pm 3230$ & $0.79 \pm 0.28$ & -- \\
    Ethenediol & HOCHCHOH & Water & $8840 \pm 3240$ & $0.76 \pm 0.28$ & -- \\
    Ethanolamine & HOCH$_{2}$CH$_{2}$NH$_{2}$ & Carbon & 8550 $\pm$ 2250 & 0.74 $\pm$ 0.19  & -- \\
    Ethanolamine & HOCH$_{2}$CH$_{2}$NH$_{2}$ & Water & 8380 $\pm$ 2260 & 0.72 $\pm$ 0.19 & -- \\
    Allenyl acetylene & H$_{2}$CCCHCCH & Carbon & 4360 $\pm$ \phantom{i} 370 & $0.37 \pm 0.03$ & -- \\ 
    Allenyl acetylene & H$_{2}$CCCHCCH & Water & 5120 $\pm$ \phantom{i} 320 & $0.44 \pm 0.03$ & -- \\ 
    Propargyl cyanide$^{\dagger}$ & HCCCH$_{2}$CN & Carbon & 6840 $\pm$ \phantom{i} 650 & $0.59 \pm 0.05$ & \multirow{2}*{18750}$^{c}$ \\ 
    Propargyl cyanide$^{\dagger}$ & HCCCH$_{2}$CN & Water & 9520 $\pm$ \phantom{i} 310 & $0.82 \pm 0.03$ &  \\ 
    Cyanoallene$^{\dagger}$ & CH$_{2}$CCHCN & Carbon & 6840 $\pm$ \phantom{i} 650 & $0.59 \pm 0.06$ & -- \\  
    Cyanoallene$^{\dagger}$ & CH$_{2}$CCHCN & Water & 9520 $\pm$ \phantom{i} 310 & $0.82 \pm 0.03$ & -- \\  
    Cyanopropyne$^{\dagger}$ & CH$_{3}$C$_{3}$N & Carbon & 6840 $\pm$ \phantom{i} 650 & $0.59 \pm 0.06$ & -- \\
    Cyanopropyne$^{\dagger}$ & CH$_{3}$C$_{3}$N & Water & 9520 $\pm$ \phantom{i} 310 & $0.82 \pm 0.03$ & -- \\
    n-Propylcyanide$^{\ddagger}$ & CH$_{3}$CH$_{2}$CH$_{2}$CN & Carbon & 7320 $\pm$ \phantom{i} 850 & $0.63 \pm 0.07$ & \multirow{2}*{21350}$^{c}$ \\
    n-Propylcyanide$^{\ddagger}$ & CH$_{3}$CH$_{2}$CH$_{2}$CN & Water & 9650 $\pm$ \phantom{i} 680 & $0.83 \pm 0.06$ &  \\
    i-Propylcyanide$^{\ddagger}$ & CH$_{3}$CH(CN)CH$_{3}$ & Carbon & 7320 $\pm$ \phantom{i} 850 & $0.63 \pm 0.08$ & -- \\
    i-Propylcyanide$^{\ddagger}$ & CH$_{3}$CH(CN)CH$_{3}$ & Water & 9650 $\pm$ \phantom{i} 680 & $0.83 \pm 0.07$ & -- \\
    Hydroxyacetone & CH$_{3}$C(O)CH$_2$OH & Carbon & $6560 \pm 3800$ & $0.57 \pm 0.33$ & -- \\
    Hydroxyacetone & CH$_{3}$C(O)CH$_2$OH & Water & $6510 \pm 3660$ & $0.56 \pm 0.31$ & -- \\
    Cyanovinylacetylene* & HCCCHCHCN & Carbon & 7430 $\pm$ \phantom{i} 860 & $0.63 \pm 0.07$ & \multirow{2}*{22600}$^{c}$ \\ 
    Cyanovinylacetylene* & HCCCHCHCN & Water & 10850 $\pm$ \phantom{i} 340 \phantom{i} & $0.93 \pm 0.03$ &  \\ 
    Vinylcyanoacetylene* & H$_{2}$CCHC$_{3}$N & Carbon & 7430 $\pm$ \phantom{i} 860 & $0.63 \pm 0.07$ & -- \\ 
    Vinylcyanoacetylene* & H$_{2}$CCHC$_{3}$N & Water & 10850  $\pm$ \phantom{i} 340  \phantom{i}  & $0.93 \pm 0.03$ & -- \\ 
    Methylcyanodiacetylene & CH$_{3}$C$_{5}$N & Carbon & $7900 \pm 1000$ & $0.68 \pm 0.09$ & \multirow{2}*{7880}$^{c}$ \\
    Methylcyanodiacetylene & CH$_{3}$C$_{5}$N & Water & 11540 $\pm$ \phantom{i} 480 \phantom{i} & $0.99 \pm 0.04$ & \\
    Cyanoacetyleneallene & H$_{2}$CCCHC$_{3}$N & Carbon & 8120 $\pm$ 1050 & 0.70 $\pm$ 0.09 & \multirow{2}*{26750}$^{c}$ \\
    Cyanoacetyleneallene & H$_{2}$CCCHC$_{3}$N & Water & 11700 $\pm$ \phantom{i} 570 \phantom{i} & 1.01 $\pm$ 0.05 &  \\
    1-cyano-1,3-cyclopentadiene & c-C$_5$H$_5$CN & Carbon & $8260 \pm 1100$  & $0.71 \pm 0.09$ & -- \\
    1-cyano-1,3-cyclopentadiene & c-C$_5$H$_5$CN & Water & 11270 $\pm$ \phantom{i} 790 \phantom{i} & $0.97 \pm 0.07$ & -- \\
    Cyanotriacetylene & HC$_{7}$N & Carbon & $8360 \pm 1130$ & $0.72 \pm 0.10$ & \multirow{2}*{7780}$^{c}$ \\
    Cyanotriacetylene & HC$_{7}$N & Water & 12040 $\pm$ \phantom{i} 630 \phantom{i} & $1.04 \pm 0.05$ &  \\
    Cyanotetraacetylene & HC$_{9}$N & Carbon & $9490 \pm 1320$ & $0.82 \pm 0.11$ & \multirow{2}*{9380}$^{c}$ \\
    Cyanotetraacetylene & HC$_{9}$N & Water & $12420 \pm 1020$ \phantom{i} & $1.07 \pm 0.09$ &  \\
    Cyanopentaacetylene & HC$_{11}$N & Carbon & $10600 \pm 1460$ \phantom{i} & $0.91 \pm 0.12$ & \multirow{2}*{10980}$^{c}$ \\
    Cyanopentaacetylene & HC$_{11}$N & Water & $12820 \pm 1260$ \phantom{i} & $1.10 \pm 0.11$ & \\
    Indene & c-C$_9$H$_8$ & Carbon & 8930 $\pm$ \phantom{i} 230 & $0.77 \pm 0.02$ & -- \\
    Indene & c-C$_9$H$_8$ & Water & 6300 $\pm$ \phantom{i} 890  & $0.54 \pm 0.08$ & -- \\
    \hline
    \label{tab:astro_species}
    \end{tabular}\
    \tablefoot{$^{\dagger}$,$^{\ddagger}$,*Isomers with identical feature descriptions have the same BEs. Literature BE estimates are taken from $^{a}$KIDA \citep[][http://kida.astrophy.u-bordeaux.fr]{wakelam2012} ; $^{b}$\citet{quan2016}; $^{c}$\citep{shingledecker2020}, the GOTHAM collaboration, and C. Shingledecker (private communication); $^{d}$\citet{belloche2019}.
    }
    \end{table*}

In recent years, the number of newly detected molecules in the interstellar medium has skyrocketed, including many complex organic molecules and prebiotic species such as carbamide \citep[NH$_{2}$C(O)NH$_{2}$,][]{belloche2019}, propargylimine \citep[HC$_{3}$HNH,][]{bizzocchi2020}, propargyl cyanide \citep[HCCCH2CN,][]{mcguire2020}, ethanolamine \citep[HOCH$_{2}$CH$_{2}$NH$_{2}$,][]{rivilla2021}, and allenyl acetylene \citep[H$_{2}$CCCHCCH,][]{cernicharo2021}. While these detections underline the molecular complexity that is present in star-forming regions, a limited knowledge of fundamental physicochemical parameters such as reaction rate constants, photo destruction cross sections, and BEs prevents us from fully understanding how these species form, react, respond to physical conditions, and, ultimately, what their place is in the interstellar chemical factory. 
     
In this section we first employ the ML model to predict the BEs of a number of molecules that have been detected in the interstellar medium, but for which no or limited information about their BEs can be found in the literature, see Table \ref{tab:astro_species}. The features of the molecules are encoded in the same way as the training data for the ML model and presented in Table \ref{tab:astro_species_features}. Predictions of BEs for monolayer coverage are limited to two surfaces, carbon and water, for which the model shows the highest performance. For a number of species, BE estimates based on the linear addition method are available in the literature and have been used in modelling studies. These BEs are generic, meaning that they are not specific to any surface. These literature estimates are included in Table \ref{tab:astro_species} for comparison to our ML predictions.

Several observations are made for the predicted BEs. The uncertainties on the predictions vary from a few \% to up to 60\% for hydroxyacetone. This is a reflection of the training data and feature representation of the model, with lower uncertainties emerging when features of a certain molecule are better represented in the training data set. In particular for molecules with cyanide groups, the uncertainties on the predicted BEs are low, because a comparatively large number of cyanide molecules are included in the training data. For about half of the molecules their predicted BE is substantially larger on a  water surface than on a carbonaceous surface. All species that show this behaviour contain a cyanide group and therefore this trend is explained by the stronger polar interaction between --CN and the H$_{2}$O ice surface. For a couple of molecules, all of them isomers, the model cannot differentiate the BEs, because they have the same feature representation.

A comparison between ML predictions and literature estimates shows some discrepancies. The predictions and estimates of methylcyanodiacetylene and the HC$_{x}$N species on carbonaceous surfaces, as well as N-methylformamide, agree fairly well, especially within the uncertainties of the ML prediction. However, the linear addition method seems to underestimate the BE of cyanamide, albeit just within the uncertainty of the ML prediction. For most species, the linear addition method severely overestimates the BE of the molecule. In the case of propargylimine the estimated BE is more than three times larger than the ML BE prediction.

Finally, for a number of species we note that the training data does not contain many molecules that represent their features, such as the imine (-N=CH-) group for ethanimine and propargylimine. Predictions for these species will have a higher degree of uncertainty. Improvements on the predictions can be expected by increasing the size of the training data set, in particular with species containing relevant features, and by expanding the feature representation. \\

We next discuss the astrophysical implications of the ML predictions. For this, we construct a simple model for the behaviour of the predicted species during thermal heating in the interstellar medium. At the basis of this model is Eq. \ref{eq:polanyi}, which is used to determine the loss of material from a surface at a specified temperature. The BEs presented in Table \ref{tab:astro_species} are used and the pre-exponential factor is assumed to be 1$\times$10$^{18}$ s$^{-1}$ for all species. With transition state theory calculations, \citet{minissale2022} showed for a selection of interstellar molecules that their prefactors are substantially larger than the canonically assumed 1$\times$10$^{12}$ or 1$\times$10$^{13}$ s$^{-1}$ and that these values increase for molecules of larger size. This motivates the choice of $\nu$ = 1$\times$10$^{18}$ s$^{-1}$, as the listed molecules are relatively large in size, but we emphasize that it is a rough assumption and values will differ from molecule to molecule. We note that when a prefactor of 1$\times$10$^{12}$ s$^{-1}$ is used, the peak desorption temperatures of molecules increase by about 20 -- 30~K, with respect to the 1$\times$10$^{18}$ s$^{-1}$ value. For each molecule, a monolayer column density of $N$ = 1$\times$10$^{15}$ molecules cm$^{-2}$ is used.

\begin{figure*}
        \centering
        \includegraphics[width=\hsize]{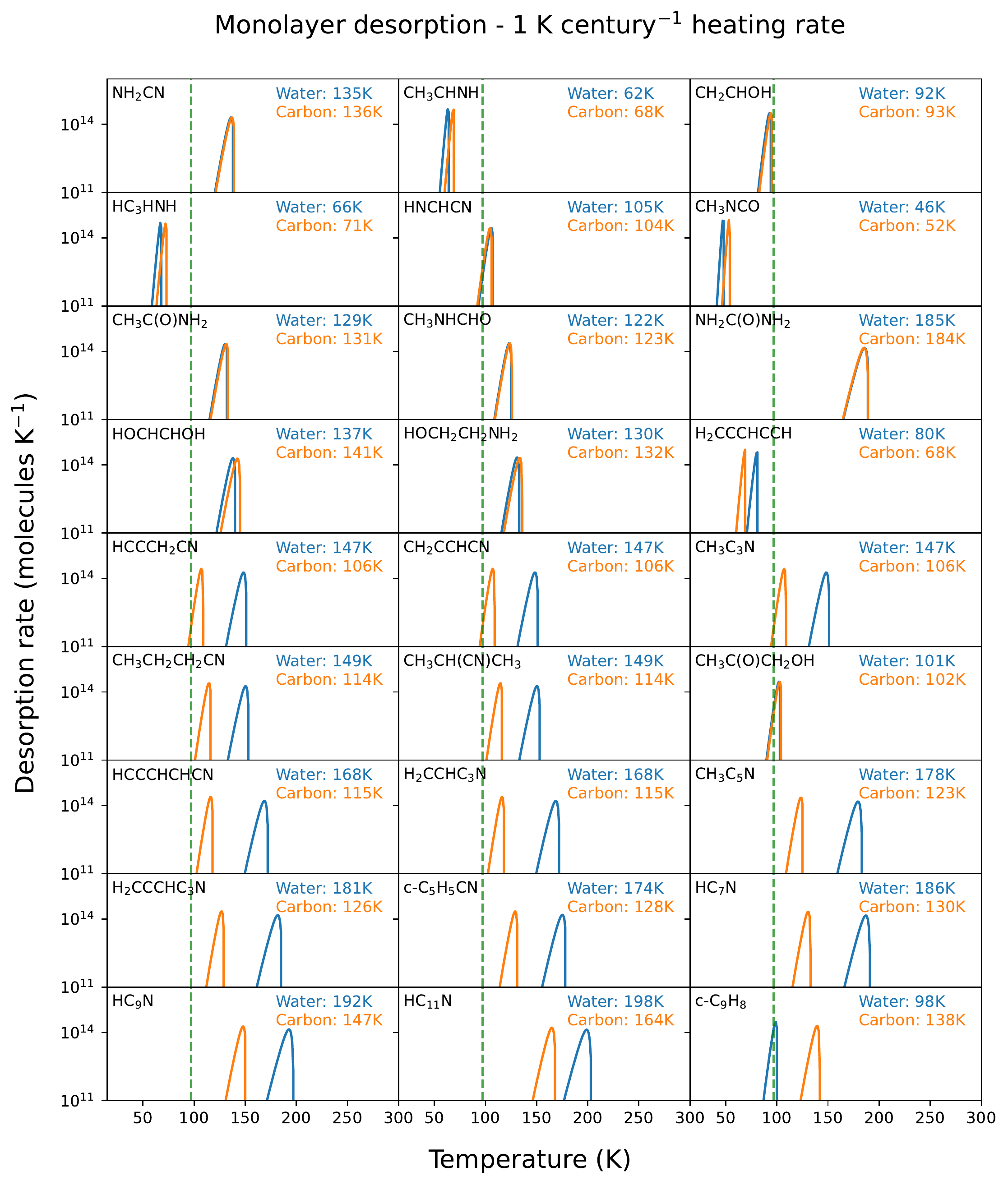}
        \caption{Desorption traces of molecules for which the BE is determined in this work. The first order desorption profiles are plotted for monolayer (1$\times$10$^{15}$ molecules cm$^{-2}$) coverage on a water ice (blue) and carbonaceous (orange) surface. The peak desorption temperatures are indicated in the top right corners. A linear heating rate of 1 K century$^{-1}$ is applied. Prefactors are assumed and set at A = 1$\times$10$^{18}$ s$^{-1}$ for all molecules. The peak desorption for water is indicated with a green dashed lines at 97 K. }
        \label{fig:des-temp}
   \end{figure*}

\begin{figure*}
        \centering
        \includegraphics[width=\hsize]{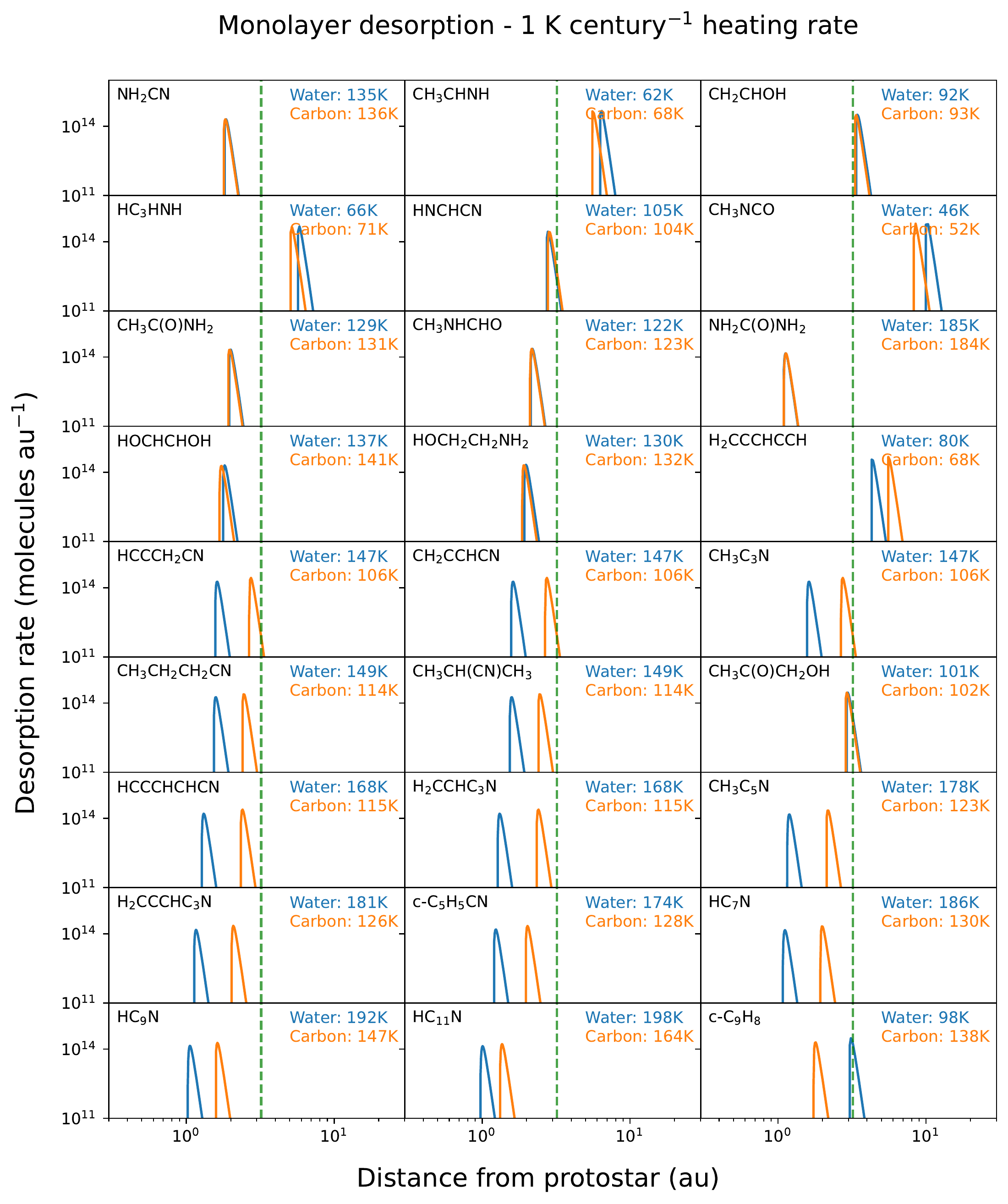}
        \caption{Same as Fig. \ref{fig:des-temp}, except that the desorption trace is plotted against a median disk temperature profile as derived by \citet{andrews2007}, see main text for more details. Shorter distances are closer to the protostar and thus correspond to higher temperatures. The peak desorption for water is indicated with a green dashed lines at 3.2 au.}
        \label{fig:des-dis}
   \end{figure*}

Two model results are presented. Figure \ref{fig:des-temp} shows the desorption profiles of the molecules versus temperature for a linear heating rate of 1 K century$^{-1}$. Figure \ref{fig:des-dis} shows the same desorption profiles, but plotted against the distance from a protostar based on the following equation:

\begin{equation}
T(r) = 200 \times r^{-0.62},
\label{eq:profile}
\end{equation}

with $r$ being the radius from the protostar in au. Equation \ref{eq:profile} is derived by \citet{andrews2007} by averaging the observed disk temperature profiles of a sample of protoplanetary disks in the Taurus-Auriga and Ophiuchus-Scorpius star forming regions. From Fig. \ref{fig:des-temp} and \ref{fig:des-dis} an indication can be given of the location of the snowlines of the molecules presented in Table \ref{tab:astro_species}, that is, the radius from a central protostar where volatile molecules sublimate or freeze out \citep{oberg2021}. In both figures the peak desorption temperature (97 K) or location (3.2 au) of water are indicated, which are determined for a monolayer (1$\times$10$^{15}$ molecules cm$^{-2}$) H$_{2}$O coverage on HOPG with $E_{\rm bin}$ = 5792 K and $\nu$ = 4.96$\times$10$^{15}$ s$^{-1}$ \citep{minissale2022}. 

The plots display a large variation in peak desorption temperatures of the predicted molecules. For some species, such as CH$_{3}$CHNH, CH$_{3}$NCO, and H$_{2}$CCCHCCH, peak desorption coincides with or is lower than that of water. While in this work the binding of molecules to a water surface is considered, it is reasonable to assume that many of these species will in fact be mixed in water ice due to the large H$_{2}$O abundance in the ISM \citep{boogert2015}. Consequently, it is to be expected that these molecule will mostly co-desorb \citep{burke2010} with water when this species sublimates. Snowlines of molecules with $E_{\rm bin,x}$ $\leq$ $E_{\rm bin,water}$ therefore more likely coincide with that of water. Co-desorption with other bulk ice components, such as CO is in principle possible, but laboratory evidence generally indicates that molecules with a significantly higher BE than the bulk medium do not co-desorb \citep[e.g.][]{ligterink2018c}. The species considered in this work, which all have $E_{\rm bin,x}$ $\geq$ $E_{\rm bin,CO}$, are therefore unlikely to co-desorb with CO. 

Many species, like CH$_{2}$CCHCN, HCCCHCHCN, and CH$_{3}$C$_{5}$N have a high BE to water surfaces and show desorption traces at much higher temperatures than the peak desorption temperature of water itself. Since at these temperatures water ice has sublimated from grain surfaces and is no longer a binding medium, desorption should instead occur from a surface that is probably made out of silicates, carbonaceous species, or organic residue. Taking this into account, it seems that many of these molecules will in fact desorb quite close to the water snow line, based on their carbon surface BEs. Only a handfull of molecules desorb at temperatures considerably above that of the water snow line, namely NH$_{2}$CN, NH$_{2}$C(O)NH$_{2}$, the HC$_{x}$N species, and indene. From this visualization it becomes clear that various subgroups of these molecules will end up in distinctly different regions of the planet-forming disk as either gas or ice. 

\section{Conclusions}
    In this work, an ML model based on Gaussian Process Regression is created and trained to predict BEs of molecules, specifically those of astrochemical relevance. The BEs determined from laboratory experiments are collected, categorized by their features (e.g., mono- or multilayer coverage, binding surface, functional groups, valence electrons, H-bond acceptors and donors), and used as training data for the model. The performance of the model is assessed with five-fold and leave-one-molecule-out cross validation. A root mean square error of 892~K and 580~K are found, for the mono- and multilayer model, respectively. For individual molecules the deviation between model predicted and literature BEs is found to be within $\pm$20\%. We note that sufficient training data and accurate feature representation are essential to predict BEs. Molecules for which features are not well described or insufficient training data points are available will generally have larger uncertainties on their predictions. 
    
    The validated model is used to predict the BEs on a water and carbonaceous surface of twenty one molecules that have been detected in recent years in the interstellar medium, but for which no or limited experimental information about their BEs is available. The lowest BE of 2990~K is predicted for methyl isocyanate (CH$_{3}$NCO) on a water surface, while the highest BE of 12820~K is predicted for cyanopentacetylene (HC$_{11}$N) bound to a water surface. Uncertainties on the predictions range from just a few percent to about 60\% for hydroxyacetone (CH$_{3}$C(O)CH$_{2}$OH), which is presumably a reflection of the lack of training data and feature representation for these molecules. The surface can have a pronounced effect on the predicted BE, showing differences of several 1000's~K for some molecules. Finally, a comparison between the ML model predictions and the in the field of astrochemistry widely used linear addition method to predict BEs is presented. We find that the linear addition methods generally overpredicts BEs, in some case by more than a factor of two. 
    
    The newly predicted BEs are put into context of interstellar environments with a simple model that shows their desorption profile with respect to a 1 K century$^{-1}$ temperature ramp and a protoplanetary disk temperature profile. From this simple model, the locations of the snowlines of these molecules are determined. Most of them will roughly coincide with the water snowline, but those of cyanamide (NH$_{2}$CN), urea/carbamide (NH$_{2}$C(O)NH$_{2}$), and the cyanoacetylenes (HC$_{x}$N) are located at much higher temperatures or closer to the protostar.
    
    This work demonstrates that ML can be employed to accurately and rapidly predict BEs of molecules. 
    The approach taken here is based on experimental training data, but we note that ML models can also be trained on BEs obtained from quantum chemical calculations, as already pursued intensively in the heterogeneous catalysis community \citep[e.g.,][]{Gu2020,Fung2021,andersen2021adsorption}. In that connection, a natural extension of this work could be to also take into account BE distributions on amorphous and highly anisotropic surfaces, as this distribution is often readily available from quantum chemical calculations \citep[e.g.,][]{tinacci2022, ferrero2020, duflot2021}. 
    Overall, we believe that the work presented here could pave the way for a stronger collaboration between the communities working on quantum chemical calculations of BEs, laboratory experiments and ML, as the various approaches complement each other. The here predicted BEs will find general use in the modelling of astrochemical and planet-forming environments, while more detailed BE distributions would be critical to more specific modelling such as the reactivity of molecules at dust grains at low temperatures.

\begin{acknowledgements}
The authors thank C.N. Shingledecker for providing BE estimates of molecules used in several chemical modelling studies. N.F.W.L. acknowledges funding by the Swiss National Science Foundation (SNSF) under Ambizione grant 193453. M.A.\ acknowledges funding from the European Union’s Horizon 2020 research and innovation programme under the Marie Sk\l{}odowska-Curie grant agreement No 754513, the Aarhus University Research Foundation, the Danish National Research Foundation through the Center of Excellence 'InterCat' (Grant agreement no.: DNRF150) and VILLUM FONDEN (grant no.\ 37381). 
\end{acknowledgements}

        \begin{table*}
    \caption{Comparison between predicted and observed values of BE obtained from the literature measured in eV and rounded to nearest 0.01.}
    \centering
    \begin{tabular}{l l l c c r}
    \hline \hline
    Name & Molecule & Surface or Coverage & Prediction & Observation & Deviation \\
    & & & (eV) & (eV) & \\
    \hline
    Acetone & $\mathrm{CH}_3 \mathrm{C(O)CH}_3$ & Water &  $0.39 \pm 0.04$ & 0.40 $\pm$ 0.02 & $-3.6$ \%  \\
    Acetonitrile & $\mathrm{CH}_3 \mathrm{CN}$ & Metal &  $0.58 \pm 0.08$  & 0.48 $\pm$ 0.03 & $21$ \%  \\
    Allyl alcohol & $\mathrm{C}_3 \mathrm{H}_5 \mathrm{OH}$ & Metal & 0.61 $\pm$  0.15  & 0.52$^{\dagger}$ \phantom{iiiiiii} & $18$ \%  \\
    Ammonia & $\mathrm{NH}_3$ & Carbon  &  $0.25 \pm 0.15$  & 0.26 $\pm$ 0.02 & $-4.0$ \%  \\
    Methane & $\mathrm{CH}_4$ & Carbon & 0.16 $\pm$ 0.04 & 0.15 $\pm$ 0.01& $10$ \%  \\
    Methyl formate & $\mathrm{CH}_3 \mathrm{OCHO}$ & Water &  $0.46 \pm 0.04$  & 0.39 $\pm$ 0.05 & $19$ \%  \\
    Nonacosane & $\mathrm{C}_{29} \mathrm{H}_{60}$ & Carbon & $2.04 \pm 0.02$ & 2.04 $\pm$ 0.00 & $0.0$ \%   \\
    Thymine & $\mathrm{C}_5 \mathrm{H}_6 \mathrm{N}_2 \mathrm{O}_2$ & Metal &  $1.01 \pm 0.22$  & 1.11 $\pm$ 0.02 & $-9.6$ \%  \\
    
    \hline
    Acetonitrile & $\mathrm{CH}_3 \mathrm{CN}$ & Multilayer &  $0.35 \pm 0.04$  & 0.42 $\pm$ 0.04 & $-16$ \%  \\
    Ammonia & $\mathrm{NH}_3$ & Multilayer &  $0.26 \pm 0.10$  & 0.26 $\pm$ 0.01 & $9.9$ \%  \\
    Methyl formate & $\mathrm{CH}_3 \mathrm{OCHO}$ & Multilayer & $0.43 \pm 0.09$  & 0.35 $\pm$ 0.02 & $10$ \%  \\
    \hline
    \label{tab:predictnewEV}
    \end{tabular}\
    \tablefoot{$^{\dagger}$Literature study does not present uncertainty of allyl alcohol BE measurement.}
    \end{table*}

\bibliographystyle{aa}
\bibliography{bib.bib}

\begin{appendix}

\section{Cross validation}
\label{ap:cross}

\begin{figure}
        \centering
        \includegraphics[width=\hsize]{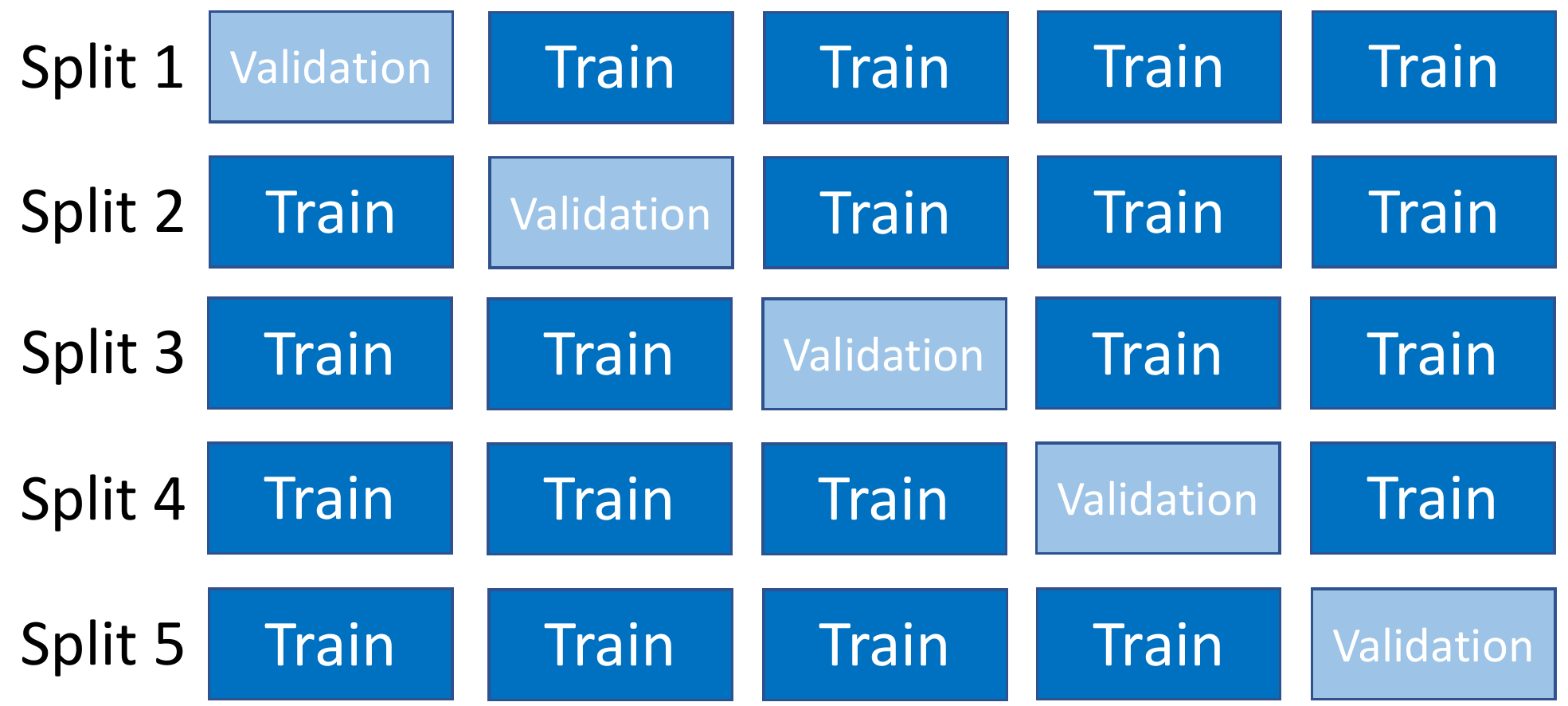}
        \caption{Schematic overview of 5-fold cross validation}
        \label{crossval}
    \end{figure}

\begin{figure}
        \centering
        \includegraphics[width=\hsize]{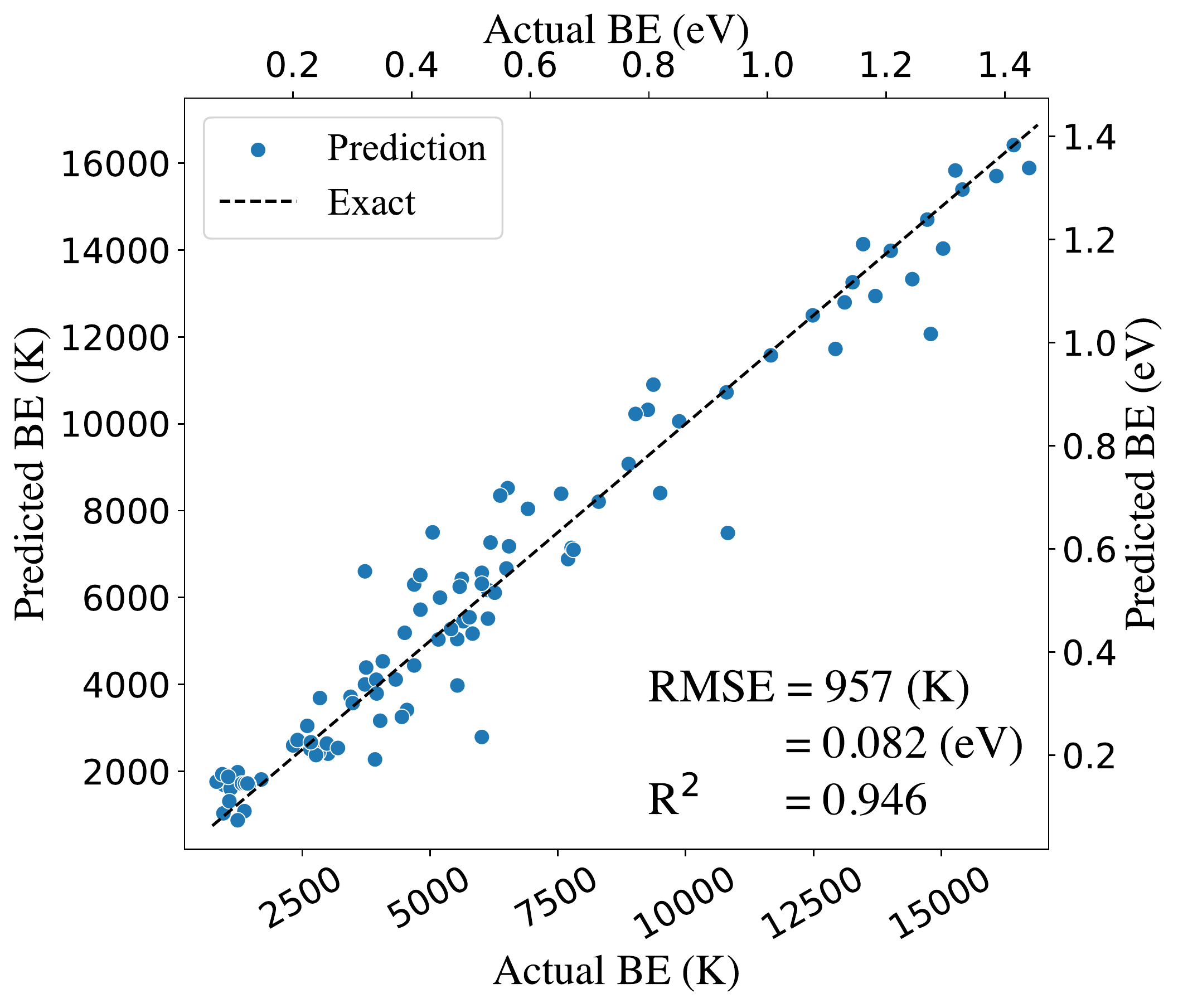}
        \caption{Parity plot for the monolayer case and including only BEs up to 17000~K.}
        \label{PPlow}
    \end{figure}

Figure \ref{crossval} schematically displays how the five-fold cross validation is set up. Figure \ref{PPlow} depicts the parity plot when only BEs up to 17000~K are included.

\section{Literature data and molecular features}
\label{ap:lit_data}

Table \ref{tab:molecule_features} shows the most important features of molecules that are used as training data in this work. Tables \ref{tab:molecule_features} and \ref{tab:multilayer_data} present the literature data used to train the ML model on\footnote{Electronic versions of the data files, along with Python scripts for producing the results, can be found in the Github repository \href{https://github.com/TorbenVilladsen/Predicting-Binding-Energies-of-Astrochemically-Relevant-Molecules-via-Machine-Learning.git}{here.}}.

\onecolumn

\begin{landscape}
\begin{longtable}{llcc|cccccccc|cccc} 
\caption{Features of molecules used for ML}\\
\hline\hline  
Molecule 	&	Molecule	&	Mass	&	atoms	&	-OH	&	-C(O)-	&	-COOH	&	-C(O)O-	&	-O-	&	-NH$_{2}$	&	-CN	&	-N-C(O)-	&	Valence 	&	H-bond 	&	H-bond 	&	TPSA	\\										
name	&	formula	&		&		&		&		&		&		&		&		&		&		&	electrons	&	acceptor	&	donor	&	\\
&	&	amu	&	\#	&	\#	&	\#	&	\#	&	\#	&	\#	&	\#	&	\#	&	\#	&	\#	&	\#	&	\#	&	Å$^2$	\\		
\hline
\endfirsthead
\multicolumn{6}{c}%
{\tablename\ \thetable\ -- \textit{Continued from previous page}} \\
\hline
Molecule 	&	Molecule	&	Mass	&	atoms	&	-OH	&	-C(O)-	&	-COOH	&	-C(O)O-	&	-O-	&	-NH$_{2}$	&	-CN	&	-N-C(O)-	&	Valence 	&	H-bond 	&	H-bond 	&	TPSA	\\										
name	&	formula	&		&		&		&		&		&		&		&		&		&		&	electrons	&	acceptor	&	donor	&	\\
&	&	amu	&	\#	&	\#	&	\#	&	\#	&	\#	&	\#	&	\#	&	\#	&	\#	&	\#	&	\#	&	\#	&	Å$^2$	\\		
\hline
\endhead
\hline \multicolumn{6}{r}{\textit{Continued on next page}} \\
\endfoot
\hline
\endlastfoot
\hline
Methane	&	CH$_{4}$	&	16	&	5	&	--	&	--	&	--	&	--	&	--	&	--	&	--	&	--	&	8	&	0	&	0	&	0	\\
Ammonia	&	NH$_{3}$	&	17	&	4	&	--	&	--	&	--	&	--	&	--	&	--	&	--	&	--	&	8	&	1	&	1	&	35	\\
Water	&	H$_{2}$O	&	18	&	3	&	1	&	--	&	--	&	--	&	--	&	--	&	--	&	--	&	8	&	0	&	0	&	32	\\
Acetylene	&	C$_{2}$H$_{2}$	&	26	&	4	&	--	&	--	&	--	&	--	&	--	&	--	&	--	&	--	&	10	&	0	&	0	&	0	\\
Hydrogen cyanide	&	HCN	&	27	&	3	&	--	&	--	&	--	&	--	&	--	&	--	&	1	&	--	&	10	&	1	&	0	&	24	\\
Carbon monoxide	&	CO	&	28	&	2	&	--	&	--	&	--	&	--	&	--	&	--	&	--	&	--	&	12	&	1	&	0	&	17	\\
Ethylene	&	C$_{2}$H$_{4}$	&	28	&	6	&	--	&	--	&	--	&	--	&	--	&	--	&	--	&	--	&	12	&	0	&	0	&	0	\\
Ethylene glycole	&	(CH$_2$OH)$_2$	&	62	&	10	&	2	&	--	&	--	&	--	&	--	&	--	&	--	&	--	&	12	&	2	&	2	&	40	\\
\end{longtable}
\tablefoot{We note that this table only lists some of the most significant features of the molecules, but does not provide the full feature list for training.}
\end{landscape}

\begin{landscape}
\begin{longtable}{llllccl} 
\caption{BEs of molecules at monolayer coverage}\\
\hline\hline  
Name	&	Formula	&	Surface	&	Simplified Surface 	&	$E_{\rm bin}$		&	$E_{\rm bin}$			&	Reference	\\
    &   &   &   &   (K)     & (eV) & \\
\hline
\endfirsthead
\multicolumn{6}{c}%
{\tablename\ \thetable\ -- \textit{Continued from previous page}} \\
\hline
Name	&	Formula	&	Surface	&	Simplified Surface 	&	$E_{\rm bin}$		&	$E_{\rm bin}$			&	Reference	\\
    &   &   &   &   (K)     & (eV) & \\
\hline
\endhead
\hline \multicolumn{7}{r}{\textit{Continued on next page}} \\
\endfoot
\hline
\endlastfoot
\hline 
1,1-Dichloroethane		&	CH$_{3}$CHCl2	&	HOPG	&	Carbon	&	\phantom{1}6134	$\pm$ \phantom{0}361	&	0.529	$\pm$	0.031	& \citet{ulbricht2006} \\			
1,2-Dichlorobenzene		&	C$_{6}$H$_{4}$Cl$_{2}$	&	HOPG	&	Carbon	&	\phantom{1}8299	$\pm$ \phantom{0}722	&	0.715	$\pm$	0.062	& \citet{ulbricht2006} \\			
2-Deoxyadenosine		&	C$_{10}$H1$_{3}$N$_{5}$O$_{3}$	&	Au(100)	&	Metal	&		13471	$\pm$ \phantom{0}481	&	1.161	$\pm$	0.041	& \citet{demers2002} \\			
2-Deoxycytidine		&	C$_{9}$H1$_{3}$N$_{3}$O$_{4}$	&	Au(100)	&	Metal	&		13711	$\pm$ \phantom{0}241	&	1.182	$\pm$	0.021	& \citet{demers2002} \\			
2-Deoxyguanosine		&	C$_{10}$H1$_{3}$N$_{5}$O$_{4}$	&	Au(100)	&	Metal	&		14433	$\pm$ \phantom{0}241	&	1.244	$\pm$	0.021	& \citet{demers2002} \\			
Acetaldehyde		&	CH$_{3}$CHO	&	Ni	&	Metal	& \phantom{1}2847	$\pm$ \phantom{00}30	&	0.245	$\pm$	0.003	& \citet{corazzi2021} \\			
Acetaldehyde		&	CH$_{3}$CHO	&	ASW	&	Water	& \phantom{1}4330	$\pm$ \phantom{0}217	&	0.373	$\pm$	0.019	& \citet{lasne2012} \\			
Acetone		&	CH$_{3}$COCH$_{3}$	&	ASW	&	Water	& \phantom{1}4691	$\pm$ \phantom{0}241	&	0.404	$\pm$	0.021	& \citet{lasne2012} \\
\end{longtable}
\tablefoot{$^{\dagger}$Species with degenerate entries for which the average of the BEs is used in the ML model. $^{a}$Based on \citet{paserba2001a,paserba2001b,gellman2002}.}
\end{landscape}

\begin{landscape}
\begin{longtable}{lllccl} 
\caption{BEs of molecules at multilayer coverage}\\
\hline\hline  
Name	&	Formula	&	Surface	&	$E_{\rm bin}$ 	&	$E_{\rm bin}$			&	Reference	\\
    &   &   &   (K)     & (eV) & \\
\hline
\endfirsthead
\multicolumn{6}{c}%
{\tablename\ \thetable\ -- \textit{Continued from previous page}} \\
\hline
Name	&	Formula	&	Surface	&	$E_{\rm bin}$	&	$E_{\rm bin}$	&	Reference	\\
    &   &   &   (K)     & (eV) & \\
\hline
\endhead
\hline \multicolumn{6}{r}{\textit{Continued on next page}} \\
\endfoot
\hline
\endlastfoot
\hline 
1,1-Dichloroethane		&	CH$_{3}$CHCl2	&	HOPG	&		5292	$\pm$	361	&	0.456	$\pm$	0.031	& \citet{ulbricht2006} \\
1,2-Dichlorobenzene		&	C$_{6}$H$_{4}$Cl$_{2}$	&	HOPG	&	\phantom{1}6735	$\pm$ \phantom{0}601	&	0.580	$\pm$	0.052	& \citet{ulbricht2006} \\
Acetic acid		&	CH$_{3}$COOH	&	Au	&		5730	$\pm$	\phantom{0}24	&	0.494	$\pm$	0.002	& \citet{kruczkiewicz2021} \\
Acetonitrile	$^{\dagger}$	&	CH$_{3}$CN	&	amorphous silica	&		4595	$\pm$	120	&	0.396	$\pm$	0.010	& \citet{abdulgalil2013} \\
Acetonitrile	$^{\dagger}$	&	CH$_{3}$CN	&	Pt(111)	&		4876	$\pm$	361	&	0.420	$\pm$	0.031	& \citet{tylinski2020} \\
Acetonitrile	$^{\dagger}$	&	CH$_{3}$CN	&	Graphene	&		5292	$\pm$	361	&	0.456	$\pm$	0.031	& \citet{tylinski2020} \\
Acetylene		&	C$_{2}$H$_{2}$	&	CsI	&		2800	$\pm$	300	&	0.241	$\pm$	0.026	& \citet{behmard2019} \\
Acrylonitrile		&	CH$_{2}$CHCN	&	Au 	&		4215	$\pm$	\phantom{0}51	&	0.363	$\pm$	0.004	& \citet{toumi2016} \\
\end{longtable}
\end{landscape}

\twocolumn

\section{Astrochemically relevant species}
\label{ap:astro_BE}

Table \ref{tab:astro_species_features} presents the features of the astrochemically relevant molecules for which BEs are predicted in this work. 

\onecolumn

\begin{landscape}
\begin{longtable}{llc|cccccccc|cccc} 
\caption{Features of astrochemically relevant molecules used for ML}\\
\hline\hline  
Molecule 	&	Molecule	&	Mass	&	-OH	&	-C(O)-	&	-COOH	&	-C(O)O-	&	-O-	&	-NH$_{2}$	&	-CN	&	-N-C(O)-	&	Valence 	&	H-bond 	&	H-bond 	&	TPSA	\\
name	&	formula	&		&		&		&		&		&		&		&		&		&	electrons	&	acceptor	&	donor	&		\\
	&		&	amu	&	\#	&	\#	&	\#	&	\#	&	\#	&	\#	&	\#	&	\#	&	\#	&	\#	&	\#	&	Å$^2$	\\
\hline
\endfirsthead
\multicolumn{6}{c}%
{\tablename\ \thetable\ -- \textit{Continued from previous page}} \\
\hline
Molecule 	&	Molecule	&	Mass	&	-OH	&	-C(O)-	&	-COOH	&	-C(O)O-	&	-O-	&	-NH$_{2}$	&	-CN	&	-N-C(O)-	&	Valence 	&	H-bond 	&	H-bond 	&	TPSA	\\
name	&	formula	&		&		&		&		&		&		&		&		&		&	electrons	&	acceptor	&	donor	&		\\
	&		&	amu	&	\#	&	\#	&	\#	&	\#	&	\#	&	\#	&	\#	&	\#	&	\#	&	\#	&	\#	&	Å$^2$	\\
\hline
\endhead
\hline \multicolumn{6}{r}{\textit{Continued on next page}} \\
\endfoot
\hline
\endlastfoot
\hline
Cyanamide	&	NH$_{2}$CN	&	42	&	--	&	--	&	--	&	--	&	--	&	1	&	1	&	--	&	16	&	2	&	1	&	50	\\
Ethanimine	&	CH$_{3}$CHNH	&	43	&	--	&	--	&	--	&	--	&	--	&	--	&	--	&	--	&	20	&	1	&	1	&	26	\\
Vinylalcohol	&	CH$_{2}$CHOH	&	44	&	1	&	--	&	--	&	--	&	--	&	--	&	--	&	--	&	18	&	1	&	1	&	20	\\
Propargylimine	&	HC$_{3}$HNH	&	53	&	--	&	--	&	--	&	--	&	--	&	--	&	--	&	--	&	20	&	1	&	1	&	24	\\
Cyanomethanimine	&	HNCHCN	&	54	&	--	&	--	&	--	&	--	&	--	&	--	&	1	&	--	&	20	&	2	&	1	&	48	\\
Methyl isocyanate	&	CH$_{3}$NCO	&	57	&	--	&	--	&	--	&	--	&	--	&	--	&	--	&	--	&	22	&	2	&	0	&	29	\\
Acetamide	&	CH$_{3}$C(O)NH$_{2}$	&	59	&	--	&	--	&	--	&	--	&	--	&	--	&	--	&	1	&	24	&	1	&	1	&	43	\\
N-Methylformamide	&	CH$_{3}$NHCHO	&	59	&	--	&	--	&	--	&	--	&	--	&	--	&	--	&	1	&	24	&	1	&	1	&	29	\\
\end{longtable}
\tablefoot{We note that this table only lists some of the most significant features of the molecules, but does not provide the full feature list for training.}
\end{landscape}

\twocolumn

\end{appendix}

\end{document}